\documentclass[cits]{PoS}

\usepackage{amsmath}
\usepackage{amssymb}
\usepackage{multirow}
\usepackage{verbatim}

\newcommand{\prd}{{\it Phys.~Rev.~D} }
\newcommand{\prl}{{\it Phys.~Rev.~Lett.} }

\newcommand{\harxold}[1]{\href{http://arxiv.org/abs/hep-lat/#1}{
[\texttt{hep-lat/#1}]} }
\newcommand{\harxnew}[1]{\href{http://arxiv.org/abs/#1}{
[\texttt{#1}]} }

\newcommand{\fortran}{\texttt{{\normalsize F}{\footnotesize ORTRAN}}~}
\newcommand{\vegas}{\texttt{{\normalsize V}{\footnotesize EGAS}}~}

\newcommand{\hippy}{\texttt{{\normalsize H}{\footnotesize I}{\normalsize
PP}{\footnotesize Y}}~}
\newcommand{\hpsrc}{\texttt{{\normalsize HP}{\footnotesize SRC}}~}
\newcommand{\mathematica}{\texttt{Mathematica}~}

\newcommand{\als}{\alpha_s}

\title{Matching heavy-light currents with NRQCD and HISQ quarks}

\ShortTitle{Matching heavy-light currents with NRQCD and HISQ quarks}

\author{\speaker{Christopher Monahan}$^{\,a}$, Christine Davies$^b$, Ron Horgan$^c$,
G.~Peter Lepage$^d$, Heechang Na$^e$ and Junko Shigemitsu$^f$\\
        \llap{$^a$}Department of Physics, College of William and Mary, Williamsburg, VA 23187, USA\\
        \llap{$^b$}SUPA, School of Physics \& Astronomy, University of Glasgow, Glasgow, G12 8QQ, UK\\
        \llap{$^c$}DAMTP, Cambridge University, Cambridge, CB3 0WA, UK\\
        \llap{$^d$}LEPP, Cornell University, Ithaca, NY 14853, USA\\
        \llap{$^e$}Argonne National Laboratory, Argonne, IL 60439, USA\\
        \llap{$^f$}Physics Department, The Ohio State University, Columbus, OH 43210, USA}
\author{HPQCD Collaboration\\
        Email: \email{cjmonahan@wm.edu}}

\abstract{We calculate the one loop renormalisation parameters for the heavy-light
axial-vector and vector currents using lattice perturbation theory. We use NonRelativistic QCD (NRQCD) heavy quarks and the Highly Improved
Staggered Quark (HISQ) action for the light quarks. We present results for
heavy-light currents with massless HISQ quarks and briefly discuss the extension to heavy-heavy
currents with massive HISQ quarks.}

\FullConference{The 30th International Symposium on Lattice Field Theory\\
June 24-29,  2012\\
Cairns, Australia}

\begin{document}

\section{Introduction}

Recent tests of the unitarity of the Cabibbo-Kobayashi-Maskawa (CKM) matrix have
indicated some tensions at the 2-3$\sigma$ level \cite{lunghi11b,ckmfitter12,utfit12}. In many cases, the constraints
on CKM unitarity are limited by the precision of the theoretical inputs, in particular the hadronic matrix elements that characterise the strong interaction dynamics of weak processes. It is therefore imperative that these
matrix elements are determined as precisely as possible.

The
HPQCD collaboration is currently undertaking a suite of precision lattice calculations of
heavy-light mesons to reduce the uncertainties associated with the theoretical inputs into CKM unitarity fits. New calculations of the decay constants $f_{B}$ and
$f_{B_s}$ using the Highly Improved Staggered Quark (HISQ) action reached a precision at the 2\%
level \cite{mcneile12a,na12}. These results are the most precise
currently available for these decay constants and were made possible by the chiral properties and reduced taste-breaking uncertainties of the HISQ action.

In \cite{na12}, the heavy-light currents were computed using HISQ light quarks and nonrelativistic QCD (NRQCD) heavy quarks. This
calculation requires matching the heavy-light axial-vector and vector
currents in the effective theory on the lattice with full QCD. In this
article we report on the one loop perturbative matching of the HISQ-NRQCD currents for massless HISQ quarks. We also discuss the extension to heavy-heavy currents.

In the next section we describe the quark and gluon actions used in
our calculation. We then review the matching formalism for heavy-light currents. In Section
\ref{sec:matchresults} we present our
results for a range of heavy quark masses and include the quark renormalisation parameters. We
discuss the extension to heavy-heavy currents in Section \ref{sec:heavyheavy} and conclude in
Section \ref{sec:summary}.

\section{Lattice Actions}

We use the Symanzik improved gluon action with tree level
coefficients \cite{luscher85a}, given by
\begin{equation}
S_G = -\frac{\beta}{3u_0^4} \sum_{x,\mu>\nu}\left[5P_{\mu\nu}
-\frac{1}{4u_0^2}\left(R_{\mu\nu}+R_{\nu\mu}\right)\right].
\end{equation}
Here $P_{\mu\nu}$ is the usual plaquette and $R_{\mu\nu}$ the six-link loop,
with $\beta = 2N_c/g^2$ and $u_0$ a tadpole improvement factor
\cite{lepage93}. Radiative improvements to the gluon
action do not contribute to our one loop matching calculation, because one loop radiative
improvement generates an ${\cal O}(\alpha_s)$ insertion in the gluon
propagator that would only contribute at higher orders.

We include a gauge-fixing term and, where possible, we confirm that gauge invariant
quantities are independent of our choice of gauge parameter by working in
both Feynman and Landau gauges, which we denote $\xi=1$ and $\xi=0$ respectively.

We use the Highly Improved
Staggered Quark (HISQ) action \cite{follana07} for the light quarks. The HISQ action has been
used successfully to simulate both $b$ and $c$
quark systems \cite{mcneile10,gregory11}. Taste-breaking discretisation errors are significantly reduced by two levels
of link fattening and ${\cal O}(a^4m_0^4)$ uncertainties are suppressed by powers of $v/c$ (where $v$ is the quark velocity) through a tuned a coefficient for the three-link ``Naik'' term \cite{follana07}. We write the action as
\begin{equation}
S_{\text{HISQ}} =
a^4\sum_x\overline{\psi}(x)\left(\gamma_{\mu}\nabla_{\mu}^{
\text{HISQ}} + m_0\right)\psi(x)\,, \quad \text{with}\quad \label{eq:nabhisq}
\nabla_{\mu}^{\text{HISQ}} = \nabla_{\mu}^{(FUF)} -
\frac{a^2}{6}(1+\epsilon)\left(\nabla_{\mu}^{(UF)}\right)^3.
\end{equation}
The superscripts indicate that the first operator,
$\nabla_{\mu}^{(FUF)}$, is built from the full HISQ-smeared links:
\begin{equation}
{\cal F}_{\mu}^{\text{HISQ}} = \left({\cal
F}_{\mu}- \sum_{\rho\neq\mu}\frac{a^2(\nabla_{\rho})^2}{2}\right){\cal U}{\cal
F}_{\mu}\,, \quad \text{where} \quad {\cal F}_{\mu}= 
\prod_{\rho\neq\mu}\left(1+\frac{a^2\nabla_{\rho}^{(2)}}{4}\right)_{
\text{symmetrised}}
\end{equation}
and ${\cal U}$ is a reunitarisation operator. The second operator in Equation \eqref{eq:nabhisq},
$\nabla_{\mu}^{(UF)}$, includes only one level of
smearing. We work with massless quarks, so we set the bare quark mass, $m_0$, and the tuning
parameter, $\epsilon$, to zero.

For the heavy quark fields, $\Psi(x,t)$, we use the NRQCD action of
\cite{gray05}, which is improved through ${\cal O}(1/M_0^2)$
and ${\cal O}(a^2)$ and includes the leading relativistic ${\cal O}(1/M_0^3)$
correction. The full NRQCD action is
\begin{align}
S_{\text{NRQCD}} = {} & 
\sum_{\mathbf{x},t}\psi^{\dagger}_{t}\psi_{t-1}
-\psi^{\dagger}_t\left(1-\frac{a\delta H}{2}\right)\left(1-
\frac{aH_0}{2n}\right)^n U_4^{\dagger}\left(1-
\frac{aH_0}{2n}\right)^n \left(1-\frac{a\delta
H}{2}\right)\psi_{t-1},
\end{align}
where $\psi^{\dagger}_{t} = \psi^{\dagger}(\mathbf{x},t)$ and
$\psi_{t-1} = \psi(\mathbf{x},t-1)$. The leading kinetic term is
given by
\begin{align}
H_0 = {} & -\frac{\Delta^{(2)}}{2aM_0}, %\nonumber \\
%\delta H = {} &
%-c_1\frac{(\Delta^{(2)})^2}{8(aM_0)^3}+c_2\frac{i}{8(aM_0)^2}
%\left(\nabla\cdot \widetilde{\mathbf{E}} - \widetilde{\mathbf{E}}\cdot
%\nabla\right) \nonumber  - c_3\frac{1}{8(aM_0)^2} \sigma\cdot
%\left(\widetilde{\nabla}\times \widetilde{\mathbf{E}} -
%\widetilde{\mathbf{E}}\times \widetilde{\nabla}\right) \nonumber \\
%{} & - c_4\frac{1}{2aM_0}\sigma\cdot
%\widetilde{\mathbf{B}}+c_5\frac{\Delta^{(4)}}{24aM_0}
%-c_6\frac{(\Delta^{(2)})^2}{16n(aM_0)^2}.
\end{align}
and $\delta H$ includes higher order improvement terms, full details of which are given in, for
example, \cite{gray05}. We use the tree level values of $c_i=1$ for all the coefficients, $c_i$, of the higher order operators in $\delta H$, and
 do not consider the effects of
radiative improvement of the NRQCD action.

\section{Matching Procedure}\label{sec:matching}

On the lattice, the heavy-light axial-vector and vector
current operators mix with higher order operators under renormalisation.
We relate the lattice and continuum currents perturbatively and extract the mixing
matrix elements at one loop. For massless HISQ quarks, the results for axial-vector
and vector currents are identical. Our strategy for the perturbative matching of heavy-light
currents follows that developed in
\cite{morningstar98} and outlined in \cite{gulez04}.% A related matching calculation
%for heavy-light currents with moving NRQCD (NRQCD formulated in a moving frame) was
%undertaken for the vector and tensor heavy-light currents in \cite{mueller11}.

We require three lattice currents to match the temporal component of the vector current through ${\cal
O}(\als,\Lambda_{\text{QCD}}/M_0,\als/(aM_0),\als\Lambda_{\text{QCD}}/M_0)$. These are
\begin{align}
J_{\mu}^{(0)} = \overline{q}(x)\Gamma_{\mu}Q(x)\,, \quad
J_{\mu}^{(1)} = 
-\frac{1}{2(aM_0)}\overline{q}(x)\Gamma_{\mu}\mathbf{\gamma}
\cdot \overrightarrow{\mathbf{\nabla}} Q(x) \,,\quad 
J_{\mu}^{(2)} = 
-\frac{1}{2(aM_0)}\overline{q}(x)\mathbf{\gamma}\cdot 
\overleftarrow{\mathbf{\nabla}}
\gamma_0 \Gamma_{\mu}Q(x). \label{eq:j2def}
\end{align}
Here the $Q$ fields are four component Dirac spinors with the upper two
components given by the two component NRQCD field and lower components
equal to zero. The $\Gamma_{\mu}$ operator represents the
vector current operator, $\Gamma_{\mu} =
\gamma_{\mu}$.

The matrix elements of the vector current in full QCD are related to those in the
effective theory via
\begin{align}
\langle V_0 \rangle = {} & \left(1+\als \rho_0^{\,(V_0)}\right)
\langle J_0^{(0)}\rangle 
+ \left(1+\als \rho_1^{\,(V_0)}\right)\langle
J_0^{(1),\,\text{sub}}\rangle  + \left(1+\als \rho_2^{\,(V_0)}\right)\langle
J_0^{(2),\,\text{sub}}\rangle; \\
\langle V_k\rangle = {} & \left(1+\als \rho_0^{\,(V_k)}\right)
\langle J_k^{(0)}\rangle + \langle J_k^{(1),\,\text{sub}} \rangle.
\end{align}
Here we have expressed the lattice currents in terms of the subtracted
currents,
\begin{equation}
J_\mu^{(i),\,\text{sub}} = J_\mu^{(i)} - \als \zeta_{i0}J_\mu^{(0)},
\end{equation}
for $i = 1,\,2$. The subtracted currents are more physical and have
improved
power law behaviour \cite{morningstar98}. Note that, for the spatial components, we
match through ${\cal O}(\als,\Lambda_{\text{QCD}}/M_0,\als/(aM_0))$.

We match at zero external quark momentum; the matching coefficients are given by
\begin{align}
\rho_0^{\,(V_0)} = {} & \frac{1}{\pi}\left(\ln(aM_0) - \frac{1}{4}\right) -
\frac{1}{2}(C_q+C_H) -
\zeta_{00}^{\,(V_0)}, \label{eq:rho0V0} \\
\rho_1^{\,(V_0)} = {} & \frac{1}{\pi}\left(\ln(aM_0) - \frac{19}{12}\right) - \frac{1}{2}(C_q+C_H) -
C_M -
\zeta_{01}^{\,(V_0)} - \zeta_{11}^{\,(V_0)}, \label{eq:rho1V0} \\
\rho_2^{\,(V_0)} = {} & \frac{4}{\pi} - \zeta_{02}^{\,(V_0)} -
\zeta_{12}^{\,(V_0)},\label{eq:rho2V0} \\
\rho_0^{\,(V_k)} = {} & \frac{1}{\pi}\left(\ln(aM_0) - \frac{11}{12}\right) - \frac{1}{2}(C_q+C_H) -
\zeta_{00}^{\,(V_k)}, \label{eq:rho0vk}
\end{align}
where the contributions from continuum QCD are given in
\cite{morningstar98,gulez04}. The renormalisation parameters $C_q$, $C_H$ and $C_M$
are the one loop HISQ wavefunction renormalisation and the NRQCD wavefunction and mass
renormalisation parameters respectively. We have
written the pole mass, which is common to both lattice and continuum
theories, in terms of the bare quark mass and must
therefore include the one loop mass renormalisation in $\rho_1$.

The $\zeta_{ij}^{\,(V_\mu)}$ in Equations \eqref{eq:rho0V0} to
\eqref{eq:rho0vk} are the one loop mixing matrix elements that arise from
the mixing of the currents. The matrix element $\zeta_{02}$ includes a term that removes an ${\cal
O}(a\als)$ discretisation error from $J_0^{(0)}$
\cite{morningstar98,gulez04}. Thus our matching procedure ensures that 
${\cal O}(\als/M_0)$ and ${\cal O}(a\als)$ corrections are made at the same
time. There is a second dimension four current operator
that is equivalent to $J_0^{(2)}$ via the equations of
motion \cite{morningstar98,gulez04}, which we include in the
determination of $\zeta_{i2}$ (for $i=1,2$).

\section{Results}\label{sec:matchresults}

We calculate the mixing matrix elements and renormalisation parameters with two independent
methods. Our first method used the automated lattice perturbation theory
routines \hippy and \hpsrc \cite{hart09}. We performed
these automated lattice perturbation theory calculations on
the Darwin cluster at the Cambridge High Performance
Computing Service and the Sporades cluster at the College of William
and Mary with routines adapted for
parallel computers using MPI (Message Passing Interface). Our second method used \mathematica and
\fortran routines to construct the appropriate Feynman integrands; these were then
evaluated with \vegas \cite{lepage80}.

We tested our code in a number of ways. We reproduced the NRQCD-AsqTad results of
\cite{gulez04} and in many cases, we established that gauge invariant
quantities, such as the mass renormalisation, were gauge parameter
independent by working in both Feynman and Landau gauges. We also confirmed that our
results exhibited the correct infrared behaviour, regulating any divergences with a gluon
mass and using subtraction functions to ensure divergences were correctly handled by
\vegas. We believe that our two methods were sufficiently independent that
agreement between these methods provides a stringent check of our results.

Many of the one loop parameters that we calculate are infrared divergent. We decompose our results into
into an infrared finite term and an infrared
divergent contribution, which we denote with a superscript $^{\text{IR}}$. Thus we write
\begin{align}
Z_q = {} & 1+\als\left(C_q^{\text{IR}}+C_q\right)+{\cal O}(\als^2)\,, \quad \quad \quad \;
C_q^{\text{IR}} = 
\frac{1}{3\pi}\left[1+\left(\xi-1\right)\right]
\log\left(a^2\lambda^2\right), \nonumber \\
Z_H = {} & 1 + \als\left(C_H^{\text{IR}}+C_H\right) + {\cal
O}(\als^2)\,, \quad \quad \quad C_H^{\text{IR}} =
\frac{1}{3\pi}\left[-2+\left(\xi-1\right)\right]
\log\left(a^2\lambda^2\right), \\
 \widetilde{\zeta}_{00} = {} & \zeta_{00} + \zeta_{00}^{\text{IR}}+ {\cal
O}(\als^2)\,,\qquad \qquad 
\qquad \; \; \zeta_{00}^{\text{IR}} =
\frac{1}{3\pi}\log(a^2\lambda^2).
\end{align}
Here $\lambda$ is the gluon mass, introduced to regulate the infrared
behaviour, and $\xi$ is the gauge parameter. We confirm that both the gluon
mass dependence and all infrared divergences cancel between the lattice and continuum one loop
coefficients. Note that the NRQCD mass renormalisation, $Z_M = 1 + \als C_M + {\cal O}(\als^2)$, is
infrared finite.

We tabulate results for the infrared finite contributions to the renormalisation parameters, mixing
matrix elements and matching parameters for the heavy-light vector current at four different heavy quark
masses in Table \ref{tab:zetaijV0}. For
the NRQCD action we present results with $c_i = 1$ and stability parameter $n=4$. We use the Landau
link definition of the tadpole improvement factor,
$u_0^{(1)}=0.7503(1)$. Only the matching coefficient $\rho_1$ has a tadpole correction coefficient.
 This correction contributes to $\zeta_{11}^{(A_0)}$ and is given by $\zeta_{11}^{u_0}=u_0^{(1)}$.
The HISQ wavefunction renormalisation is of course independent of the NRQCD
mass and we find $C_q(\xi=0) = 0.1145(1)$ and $C_q(\xi=1) = -0.3940(1)$.
\begin{comment}
\begin{table}\label{tab:renormparms}
\begin{tabular}{cc|cccc}
$C_q(\xi=0)$ & $C_q(\xi=1)$ & $aM_b$ & $C_H(\xi=0)$ & $C_H(\xi=1)$ &
$C_M$ \\
& & & & \vspace*{-8pt} \\
\hline
& & & & \vspace*{-8pt} \\
\multirow{4}{*}{-0.8183(1)} & \multirow{4}{*}{-0.0198(1)} & 2.688 & \hphantom{-}0.147(2) & -0.360(2)
& 0.262(1) \\
& & 2.650 & \hphantom{-}0.139(2) & -0.370(2) & 0.267(1) \\
& & 1.832 & -0.153(2) & -0.658(2) & 0.466(1) \\
& & 1.826 & -0.157(2) & -0.662(2) & 0.468(1) \\
\vspace*{-5pt}\\
\end{tabular}
\caption{One-loop renormalisation parameters required for one loop matching. All uncertainties are
statistical
errors arising from the numerical integration of the relevant diagrams. The HISQ wavefunction,
$C_q$,
renormalisation is independent of the NRQCD
mass.}
\end{table}

We tabulate our results for the mixing matrix elements
$\zeta_{ij}^{\,(V_0)}$ at four
different heavy quark masses in Table \ref{tab:zetaijV0}.
\end{comment}
\begin{table}\label{tab:zetaijV0}
\begin{tabular}{ccccccccc}
\vspace*{-5pt}\\
$aM_0$ & $C_H$ & $C_M$ & $\rho_0^{\,(V_0)}$ & $\rho_1^{\,(V_0)}$ &
$\rho_2^{\,(V_0)}$ & $\zeta_{10}^{\,(V_0)}$ &  $\rho_0^{\,(V_k)}$ & $\zeta_{10}^{\,(V_k)}$ \\
\vspace*{-5pt}\\
\hline
\vspace*{-5pt}\\
2.688 & -0.360(2) & 0.262 & -0.108(2) & \hphantom{-}0.012(2) & -0.712(4) & -0.1144 & -0.034(2)
& 0.0382 \\
2.650 & -0.370(2) & 0.267 & -0.111(2) & \hphantom{-}0.013(2) & -0.693(4) & -0.1157 & -0.034(2)
& 0.0386 \\
1.832 & -0.658(2) & 0.466 & -0.162(2) & -0.042(3) & -0.314(4) & -0.1593 & \hphantom{-}0.018(2)
& 0.0532 \\
1.826 & -0.662(2) & 0.468 & -0.164(2) & -0.043(3) & -0.311(4) & -0.1595 &
\hphantom{-}0.020(2)
& 0.0532 \\
\vspace*{-5pt}\\
\end{tabular}
\caption{One-loop renormalisation parameters for the
heavy-light vector current in Feynman gauge. For the $\rho_i^{\,(V_\mu)}$ the quoted
uncertainties
are the errors from each contribution added in quadrature, whilst for $C_H$, $C_M$ and the
$\zeta_{10}^{\,(V_\mu)}$ the uncertainty is the statistical error
from numerical integration. Unless otherwise indicated, the uncertainties are 1 or smaller in the
final digit.}
\end{table}

\section{Heavy-heavy Currents}\label{sec:heavyheavy}

We are currently extending our calculation to heavy-heavy currents, with
non-zero HISQ mass. Moving from massless to massive relativistic quarks complicates the
matching procedure. In the former case, quarks and antiquarks at zero
spatial momentum are indistinguishable and consequently scattering and annihilation processes give
identical results. In the massive case, however, we must distinguish between
quarks and antiquarks. Massive HISQ quarks also complicate the numerical
 integration considerably. The chief difficulty arises for the $A_0$ and $V_k$ annihilation currents,
which contain a
Coulomb singularity that must be handled with a subtraction
function. Moreover, in the automated
perturbation theory
routines, the pole in NRQCD propagator crosses the integration contour and we must 
introduce a triple contour to ensure the stability of
numerical integration
\cite{mueller11}. 

\section{Summary}\label{sec:summary}

We have determined the one loop matching coefficients and renormalisation parameters required to
match
the axial-vector and vector currents on the lattice to full QCD. We
used the massless HISQ action for the light
quarks and NRQCD for the heavy quarks and match at zero external quark momentum.

These matching coefficients are important ingredients in the determination
of heavy-light mesonic decays in lattice QCD studies \cite{na12}. Studies of the $B_s$ meson using the relativistic HISQ action for both $b$
and $s$ quarks have recently been carried out \cite{mcneile12a}. Such an
approach has the advantage that perturbative matching, which
is generally the dominant source of error in the extraction of decay
constants, is not required. Computations at the
physical $b$ quark mass are currently prohibitively expensive, however, and require an extrapolation
up to the $b$ quark mass. Furthermore, simulations of the
$B$ meson are not yet feasible, because the use of light valence quarks
and close-to-physical $b$ quark masses require large lattices with
fine lattice spacings. Thus an
effective theory approach to heavy-light systems remains the most efficient framework
for precise predictions of $f_{B_s}/f_B$ and $f_B$. Such calculations
require the perturbative matching calculation reported in this article. 

The extension to heavy-heavy currents will enable the extraction of form factors for the
$B\rightarrow D^{(\ast)}\ell\nu$ semileptonic decays and the leptonic decay of the $B_c$ meson
from lattice NRQCD computations. These calculations are underway.%In
%principle it should be possible to extend the matching procedure to two
%loops, further improving the precision of lattice QCD decay constant
%determinations. 

% Use the proper section head for acknowledgments.
\begin{acknowledgments}
The authors would like to thank Georg von Hippel 
for many helpful
discussions. This work was supported by the
DOE, the NSF and the STFC. Some of the computing was undertaken on the Darwin supercomputer at the HPCS, University of Cambridge, as part of 
the DiRAC facility jointly funded by the STFC. 
 \end{acknowledgments}

\end{document}